\documentclass[showpacs,preprintnumbers,english,amsmath,amssymb]{revtex4}
\usepackage[T1]{fontenc}
\usepackage[latin1]{inputenc}
\usepackage{graphicx}
\usepackage{amssymb}
\usepackage{verbatim}
\usepackage{epic,eepic}
\usepackage{babel}
\usepackage{color}

\begin{document}

\def\Journal#1#2#3#4{{#1}, {#2}, {#3}, {#4}.}
\def\ADEP{Advances in High Energy Physics}
\def\ANP{Adv. Nucl. Phys.}
\def\ARNPS{Ann. Rev. Nucl. Part. Sci.}
\def\CTP{Commun. Theor. Phys.}
\def\CPL{Chin. Phys. Lett.}
\def\EPJA{The European Physical Journal A}
\def\EPJC{The European Physical Journal C}
\def\IJMPA{International Journal of Modern Physics A}
\def\IJMPE{International Journal of Modern Physics E}
\def\JCHP{J. Chem. Phys.}
\def\JCP{Journal of Computational Physics}
\def\JHEP{Journal of High Energy Physics}
\def\JPCS{Journal of Physics: Conference Series}
\def\JPG{Journal of Physics G: Nuclear and Particle Physics}
\def\NATURE{Nature}
\def\NC{La Rivista del Nuovo Cimento}
\def\NCA{IL Nuovo Cimento A}
\def\NPA{Nuclear Physics A}
\def\NPB{Nuclear Physics B}
\def\NST{Nuclear Science and Techniques}
\def\PA{Physica A}
\def\PAN{Physics of Atomic Nuclei}
\def\PHY{Physics}
\def\PRA{Phys. Rev. A}
\def\PRC{Physical Review C}
\def\PRD{Physical Review D}
\def\PLA{Phys. Lett. A}
\def\PLB{Physics Letters B}
\def\PLD{Phys. Lett. D}
\def\PRL{Physical Review Letters}
\def\PL{Phys. Lett.}
\def\PREV{Phys. Rev.}
\def\PREP{\em Physics Reports}
\def\PROG{Progress in Particle and Nuclear Physics}
\def\RPP{Rep. Prog. Phys.}
\def\RDNC{Rivista del Nuovo Cimento}
\def\RMP{Rev. Mod. Phys}
\def\SCIENCE{Science}
\def\ZPA{Z. Phys. A.}

\def\ANN{Ann. Rev. Nucl. Part. Sci.}
\def\ANNAST{Ann. Rev. Astron. Astrophys.}
\def\AP{Ann. Phys}
\def\APJ{Astrophysical Journal}
\def\APJS{Astrophys. J. Suppl. Ser.}
\def\EJP{Eur. J. Phys.}
\def\LANC{Lettere Al Nuovo Cimento}
\def\NCA{Nuovo Cimento A}
\def\PHYS{Physica}
\def\NP{Nucl. Phys}
\def\MATH{J. Math. Phys.}
\def\JPAM{J. Phys. A: Math. Gen.}
\def\PRO{Prog. Theor. Phys.}

\title{Comparing the Tsallis distribution with and without thermodynamical description in p+p collisions}

\author{H. Zheng$^{1}$\footnote{Email address: zheng@lns.infn.it} and Lilin Zhu$^{2}$\footnote{Email address: zhulilin@scu.edu.cn}}
\affiliation{
$^{1}$Laboratori Nazionali del Sud, INFN, via Santa Sofia, 62, 95123 Catania, Italy;\\
$^{2}$College of Physical Science and Technology, Sichuan University, Chengdu 610064, People's Republic of China.}


\begin{abstract}
We compare two types of Tsallis distribution, i.e., with and without thermodynamical description, using the experimental data from the  STAR, PHENIX, ALICE and CMS Collaborations on the rapidity and energy dependence of the transverse momentum spectra in p+p collisions. Both of them can give us the similar fitting power to the particle spectra. We show that the Tsallis distribution with thermodynamical description gives lower temperatures than the ones without it. The extra factor $m_T$ (transverse mass) in the Tsallis distribution with thermodynamical description plays an important role in the discrepancies between the two types of Tsallis distribution. But for the heavy particles, the choice to use the $m_T$  or $E_T$ (transverse energy) in the Tsallis distribution becomes more crucial.
\end{abstract}


\maketitle

\section{Introduction}
The particle spectrum is a basic quantity directly measured in the experiments and it can reveal the information of particle production mechanism in heavy-ion collisions. Many physicists have devoted themselves to studying the particle spectra produced in the heavy-ion collisions using thermodynamical approaches, phenomenological methods, transport models {\it et al.} \cite{tat, pq, hy, vg, sena,  liuAuAu2014, khandaiflow, cleymans, azmiJPG2014,  liAuAu2013, maciej, khandai, wongprd, wong2012, wongarxiv2014, maciej1, beck, dndeta, huacluster, huapp, huaaa, cleymansplb2013}. Recently, the Tsallis distribution, which was first proposed about twenty-seven years ago as a generalization of the Boltzmann-Gibbs distribution \cite{tsallis},  has attracted many theorists' and experimentalists' attention in high energy heavy-ion collisions \cite{star2007, phenix2011, alice1, alice2, aliceS2012, cms3, cmsdata900, cmsdata7000, cms1, sena,  liuAuAu2014, khandaiflow,  cms2014,  cleymans, azmiJPG2014,  liAuAu2013, maciej, khandai, wongprd, wong2012, wongarxiv2014, maciej1}. The excellent ability to fit the spectra of identified hadrons and charged particles in a large range of $p_T$ up to 200 GeV, which covers 15 orders of magnitude, is quite impressive. This spectacular result was first shown by Wong {\it et al.}  \cite{wongprd, wong2012, wongarxiv2014, maciej1}. In refs. \cite{huapp, huaaa}, we have shown that Tsallis distribution can fit almost all the particle spectra produced in p+p, p+A and A+A at RHIC and LHC. From the phenomenological view, there may be real physics behind the prominently phenomenological work, e.g., Regge trajectory for particle classification \cite{wongbook}. We also note that there are different versions of Tsallis distribution in the literature and we classify them as Type-A, B and C to clarify the comparison in ref. \cite{huapp}. Type-A Tsallis distribution is obtained without resorting to thermodynamical description, but it has been adopted to analyze the particle spectra by STAR \cite{star2007}, PHENIX \cite{phenix2011} Collaborations at RHIC and ALICE \cite{alice1, alice2, aliceS2012}, CMS \cite{cms3} Collaborations at LHC. 
In ref. \cite{huapp}, we applied it to do the systematic analysis of identified particle spectra in p+p collisions at RHIC and LHC and proposed a cascade particle production mechanism. On the other hand, Type-B Tsallis distribution is derived by taking into account the thermodynamical consistency and is widely used by J. Cleymans and his collaborators to study the particle spectra in high-energy p+p collisions \cite{ cleymans, azmiJPG2014, cleymansplb2013}. It is also used by the other authors \cite{liAuAu2013}. Type-A and B are the most popular Tsallis distributions in the literature but they give quite different temperatures while fitting the same particle spectra, e.g., for pion, Type-A gives $T\sim 0.13$ GeV while Type-B gives $T\sim 0.075$ GeV. In this paper, we would like to systematically address the question what results in the discrepancies of the temperatures for the two types of Tsallis distribution using particle spectra in p+p collisions. The data produced in p+p collisions with different $p_T$ ranges and different rapidity cuts are collected from the experimental Collaborations at RHIC and LHC \cite{phenix2011, alice1, alice2, alice4, alice5, cms1, phenix2009, phenix2007, star2, cms2, alice3}.

The paper is organized as follows. In Sec. II, we introduce the two types of Tsallis distribution without and with thermodynamical description in our comparison. In order to make our discussion clear, we also introduce another three transient distributions which are very similar to Type-A and Type-B Tsallis distributions. In Sec. III, the results of particle spectra from the different distributions in p+p collisions and the comparisons are shown. A brief conclusion is given in the Sec. IV.

\section{Tsallis distributions}
Type-A Tsallis distribution has been widely adopted by STAR \cite{star2007}, PHENIX \cite{phenix2011} Collaborations at RHIC and ALICE \cite{alice1, alice2, aliceS2012}, CMS \cite{cms3} Collaborations at LHC
\begin{equation}
E\frac{d^3N}{dp^3} = \frac{dN}{dy} \frac{(n-1)(n-2)}{2\pi nC[nC+m(n-2)]}(1+\frac{m_T-m}{nC})^{-n}, \label{exptsallis}
\end{equation}
where $m_T=\sqrt{p_T^2+m^2}$ is the transverse mass. $m$ was used as a fitting parameter in ref. \cite{star2007}, but it represents the rest mass of the particle studied in refs. \cite{phenix2011, alice1, alice2, aliceS2012, cms3}.  $\frac{dN}{dy}$, $n$ and $C$ are fitting parameters. When $p_T\gg m$, we can ignore the $m$ in the last term in Eq. (\ref{exptsallis}) and obtain $E\frac{d^3N}{dp^3} \propto p_T^{-n}$. This result is well known because high energy particles come from hard scattering and they follow a power law distribution with $p_T$. When $p_T\ll m$ which is the non-relativistic limit, we obtain $m_T -m=\frac{p_T^2}{2m}=E_{T}^{classical}$ and $E\frac{d^3N}{dp^3} \propto e^{-\frac{E_T^{classical}}{C}}$, i.e., a thermal distribution. The parameter $C$ in Eq. (\ref{exptsallis}) plays the same role as temperature $T$. In refs. \cite{huapp, huaaa}, we have obtained the simpler form of Eq. (\ref{exptsallis})
\begin{equation}
(E\frac{d^3N}{dp^3})_{|\eta|<a} = A(1+\frac{E_T}{nT})^{-n}, \label{tsallisus}
\end{equation}
where $E_T=m_T - m$. $A$, $n$ and $T$  are free fitting parameters in Eq. (\ref{tsallisus}). We note that it has been used by CMS Collaboration \cite{cmsdata900, cmsdata7000, cmsB2012} and by Wong {\it et al.} in their recent paper \cite{wongarxiv2014}. The STAR Collaboration also applied a formula which is very close to Eq. (\ref{tsallisus}) \cite{star2010}. We adopt the Eq. (\ref{tsallisus}) in the following study.

In the framework of Tsallis statistics, the distribution function is
\begin{equation}
f(E, q) = [1+(q-1)\frac{E-\mu}{T}]^{-\frac{1}{q-1}}.
\end{equation}
Taking into account the self-consistent thermodynamical description, one has to use the effective distribution $[f(E, q)]^q$. Therefore, the Tsallis distribution is obtained  \cite{cleymans, azmiJPG2014, liAuAu2013, maciej}
\begin{eqnarray}
E\frac{d^3N}{dp^3} = gV\frac{m_T \cosh y}{(2\pi)^3} [1+(q-1)\frac{m_T\cosh y-\mu}{T}]^{-\frac{q}{q-1}}, \label{tsallisB}
\end{eqnarray}
where $g$ is the degeneracy of the particle, $V$ is the volume, $y$ is the rapidity, $\mu$ is the chemical potential, $T$ is the temperature and $q$
is the entropic factor, which measures the non-additivity of the entropy. We dubbed it as the Type-B Tsallis distribution \cite{huapp}. In Eq. (\ref{tsallisB}), there are four parameters $V, \mu, T, q$. $\mu$ was assumed to be 0 in refs. \cite{cleymans, azmiJPG2014, liAuAu2013} which is a reasonable assumption because the energy is high enough and the chemical potential is much smaller than the temperature. In the mid-rapidity $y=0$ region, Eq. (\ref{tsallisB}) is reduced to
\begin{equation}
E\frac{d^3N}{dp^3} = gV\frac{m_T}{(2\pi)^3} [1+(q-1)\frac{m_T}{T}]^{-q/(q-1)}. \label{tsallisBR}
\end{equation}
It becomes very similar to Eq. (\ref{tsallisus}), but there are some differences, e.g., $m_T$ replaces $E_T$ in the bracket and there is an extra term $m_T$ in front of the bracket. It should be pointed out that there is no direct match between $n$ and $q$ in Eqs. (\ref{tsallisus}, \ref{tsallisBR}). But we could find a connection between $n$ and $q$ in the limit at large $p_T$. When $p_T\gg m$, from Eq. (\ref{tsallisBR}), we can obtain
\begin{equation}
E\frac{d^3N}{dp^3} \propto p_T^{-\frac{1}{q-1}}. \label{tsallisBRlimit}
\end{equation}
Recalling that $E\frac{d^3N}{dp^3}\propto p_T^{-n}$ when $p_T\gg m$ from Eq. (\ref{exptsallis}), therefore the relation between $n$ and $q$ is
\begin{equation}
n = \frac{1}{q-1}. \label{nq}
\end{equation}
Another treatment to find the relation between $n$ and $q$ can be found in ref. \cite{cleymansplb2013}.

For the other Tsallis distributions in the literature, we refer them to refs. \cite{huapp, huaaa}. We noted that Type-A and Type-B Tsallis distributions can reproduce the particle spectra in p+p collisions very well but Type-B gives lower temperatures than the ones given by Type-A. In this paper, we would like to address this discrepancies between the two types of Tsallis distribution. To make our discussion clear, another three transient distributions are used to bridge the Type-A and Type-B distributions. In ref. \cite{transient1}, a Tsallis-like distribution is obtained in the framework of non-extensive statistics for the particle invariant yield at midrapidity
\begin{equation}
E\frac{d^3N}{dp^3}=A m_T [1+(q-1)\frac{m_T}{T}]^{-\frac{1}{q-1}}, \label{transeq1}
\end{equation}
where $A$, $q$ and $T$ are fitting parameters. Comparing Eq. (\ref{transeq1}) and Eq. (\ref{tsallisBR}), the only difference is the power of the distribution function, i.e., $q$ for Eq. (\ref{tsallisBR}) and $1$ for Eq. (\ref{transeq1}). We also introduce another two forms of distribution. One is
\begin{equation}
E\frac{d^3N}{dp^3}=A [1+(q-1)\frac{m_T}{T}]^{-\frac{q}{q-1}}, \label{transeq2}
\end{equation}
where we neglect the term $m_T$ outside of the bracket in Eq. (\ref{tsallisBR}) and the constants are absorbed into the parameter $A$. The other one is
\begin{equation}
E\frac{d^3N}{dp^3}=A [1+(q-1)\frac{m_T}{T}]^{-\frac{1}{q-1}}. \label{transeq3}
\end{equation}
Similar to Eq. (\ref{transeq2}), we neglect the term $m_T$ in front of the bracket in Eq. (\ref{transeq1}) but keep the rest. Using the relation Eq. (\ref{nq}), we find that Eq. (\ref{transeq2}) becomes Eq. (\ref{tsallisus}) in the limit of massless particle.

Let us put the four distributions in order of Eqs. (\ref{tsallisBR}, \ref{transeq1}, \ref{transeq3}, \ref{tsallisus}) and Eqs. (\ref{tsallisBR}, \ref{transeq2}, \ref{transeq3}, \ref{tsallisus}), one can see that any adjacent distributions have only one term different. We successfully bridge the Type-A and Type-B Tsallis distributions and are able to conduct our investigation. We need to point out that even though we use the same symbols for the parameters in the five distributions, they may have different values when fitting the experimental data. In next section, we will systematically apply the five distributions to the particle spectra in p+p collisions, similar to our previous work \cite{huapp}. But we update some experimental data which have larger $p_T$ ranges and will focus on the temperature differences among the five distributions.

\begin{figure}
\resizebox{0.45\textwidth}{!}{%
  \includegraphics{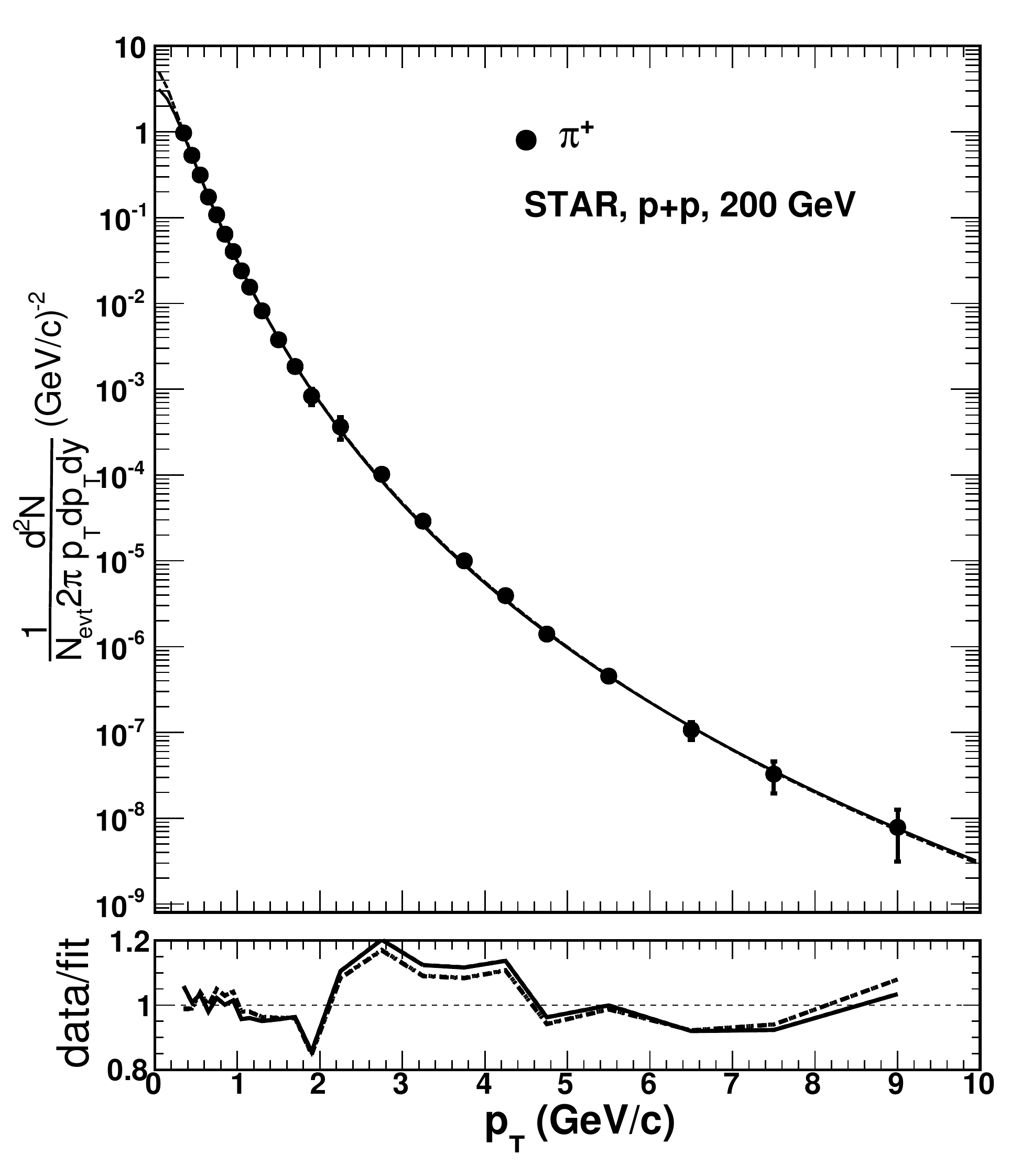}
}
\caption{Fitting results using the distributions Eqs. (\ref{tsallisBR}, \ref{transeq1}, \ref{transeq2}, \ref{transeq3}, \ref{tsallisus}) for $\pi^+$ in p+p collisions at $\sqrt{s}=200$ GeV. The solid line, dashed line, dotted line, dash-dotted line and long dash-dotted line refer to Eqs. (\ref{tsallisBR}, \ref{transeq1}, \ref{transeq2}, \ref{transeq3}, \ref{tsallisus})  respectively, but they are hardly to distinguish because they give us similar fitting power. The ratios of data/fit are shown at the bottom. Data are taken from STAR \cite{star2}. }
\label{Fig1}       
\end{figure}

\begin{figure}
\resizebox{0.45\textwidth}{!}{%
  \includegraphics{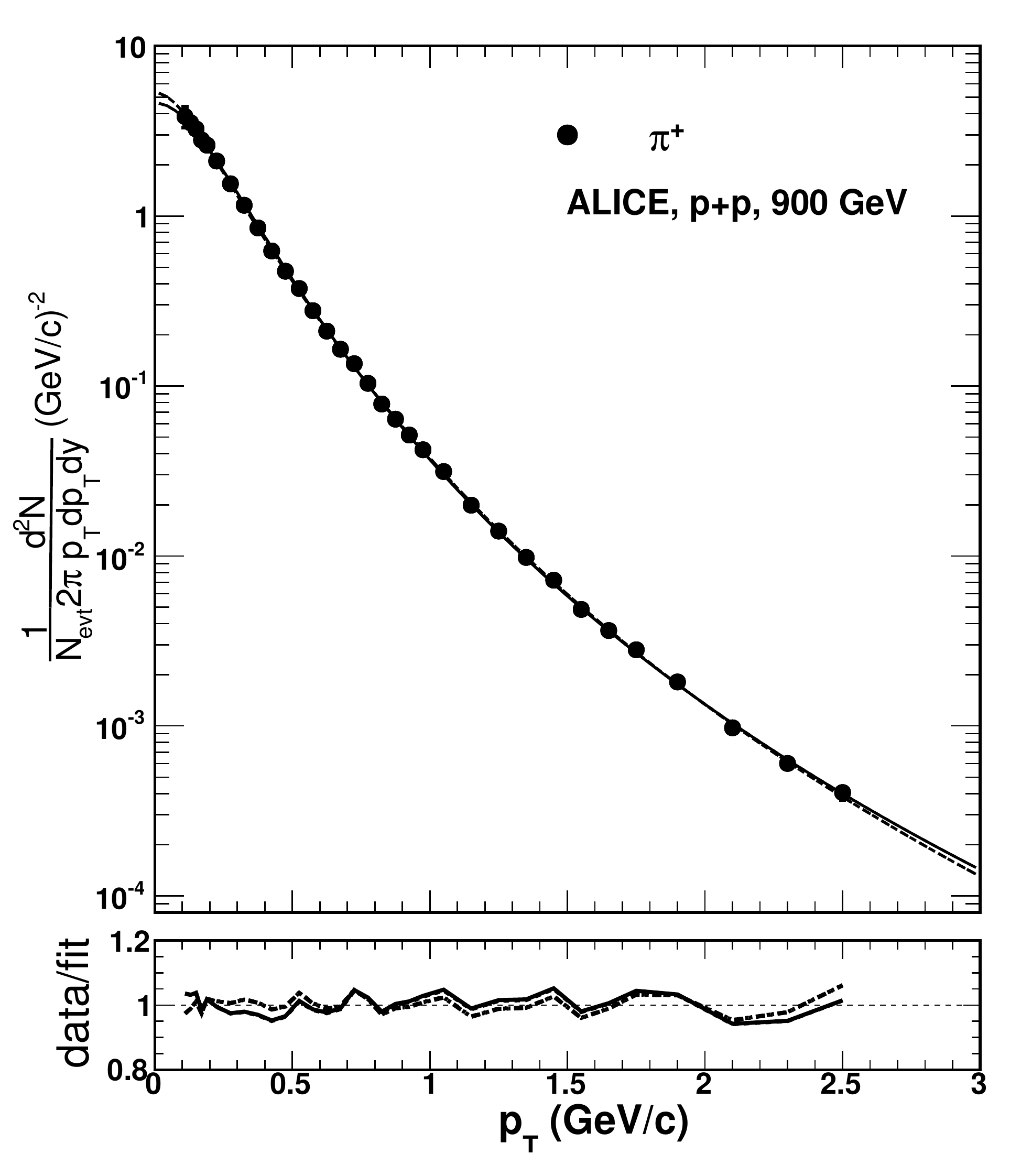}
}
\caption{Same as in Fig. \ref{Fig1} for $\pi^+$ in p+p collisions at $\sqrt{s}=900$ GeV. Data are taken from ALICE \cite{alice2}. }
\label{Fig2}       
\end{figure}

\begin{figure}
\resizebox{0.45\textwidth}{!}{%
  \includegraphics{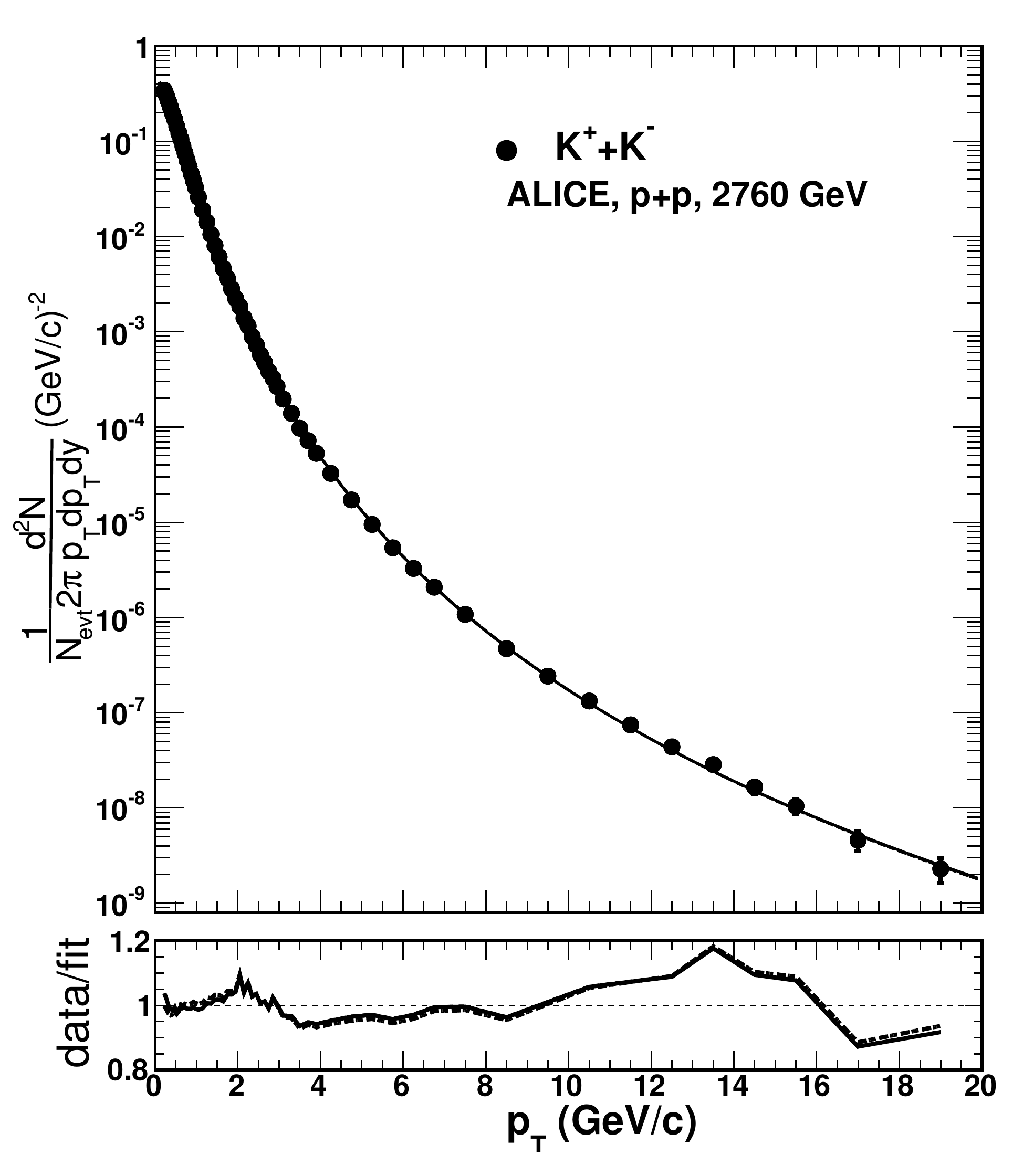}
}
\caption{Same as in Fig. \ref{Fig1} for $K^++K^-$ in p+p collisions at $\sqrt{s}=2760$ GeV. Data are taken from ALICE \cite{alice4}.}
\label{Fig3}       
\end{figure}

\begin{figure}
\resizebox{0.45\textwidth}{!}{%
  \includegraphics{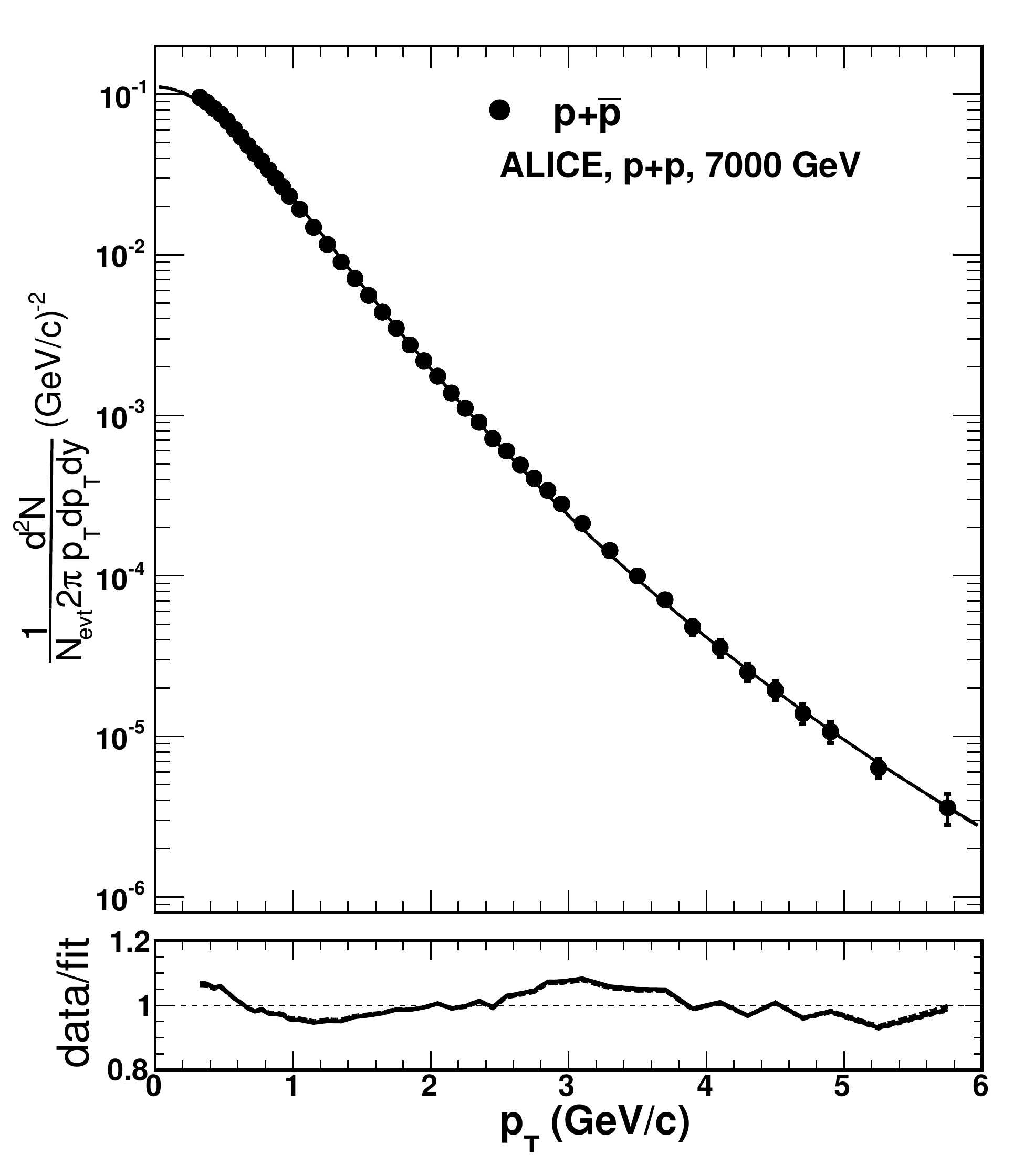}
}
\caption{Same as in Fig. \ref{Fig1} for $p+\bar{p}$ in p+p collisions at $\sqrt{s}=7000$ GeV. Data are taken from ALICE \cite{alice5}.}
\label{Fig4}       
\end{figure}

         \begin{figure*}
        \centering
        \begin{tabular}{ccc}
        \includegraphics[width=0.32\textwidth]{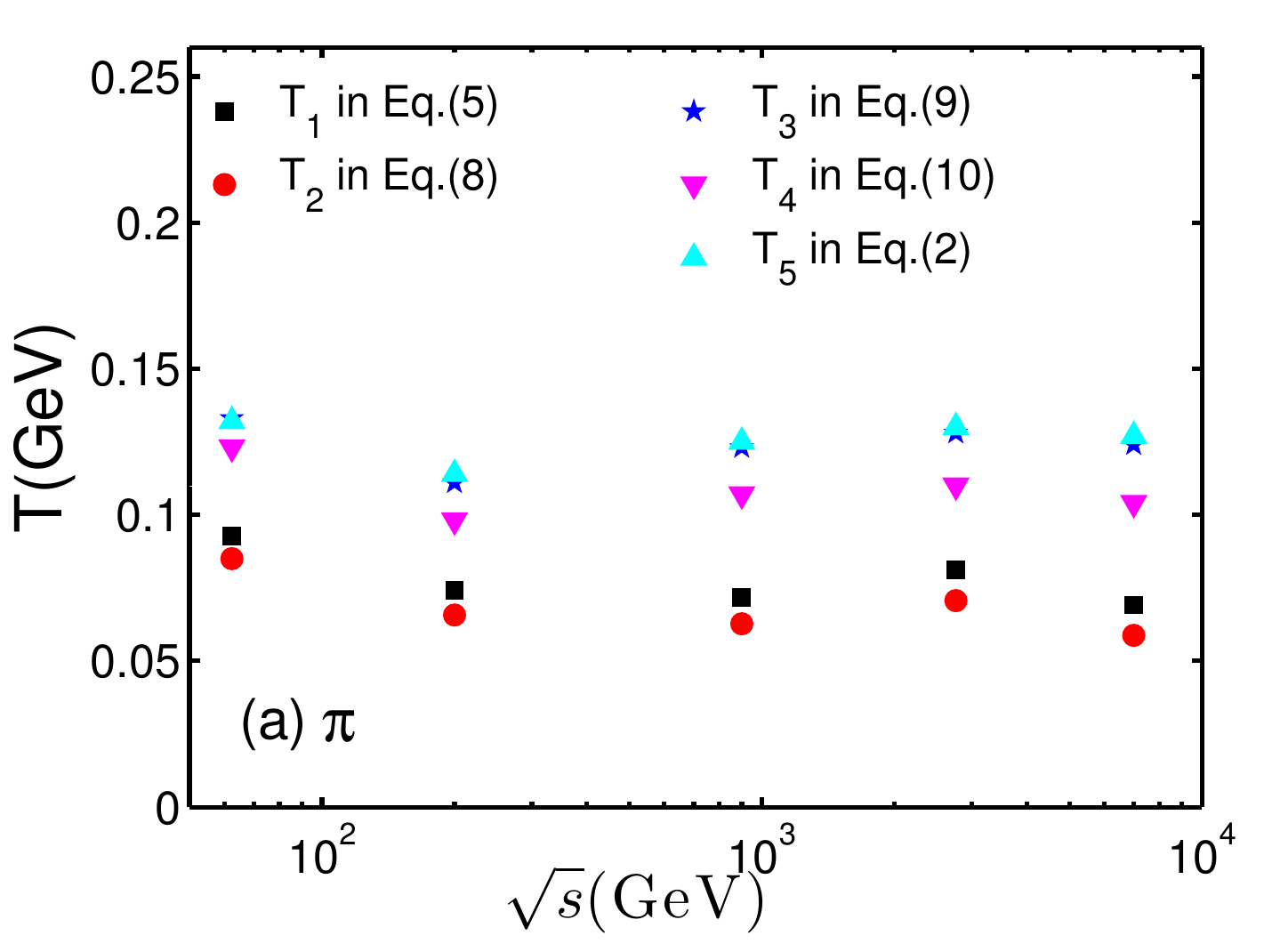}
        \includegraphics[width=0.32\textwidth]{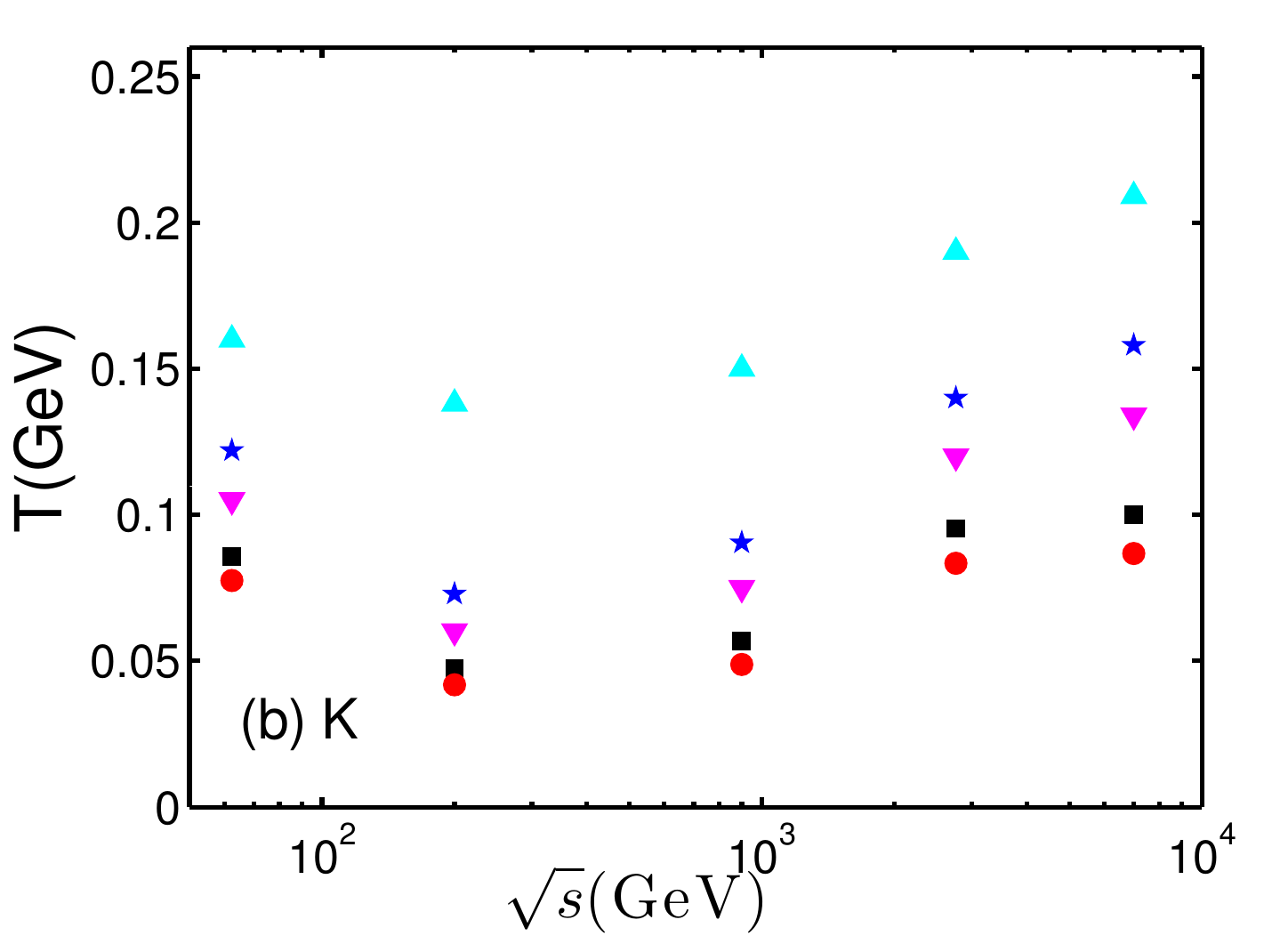}
          \includegraphics[width=0.32\textwidth]{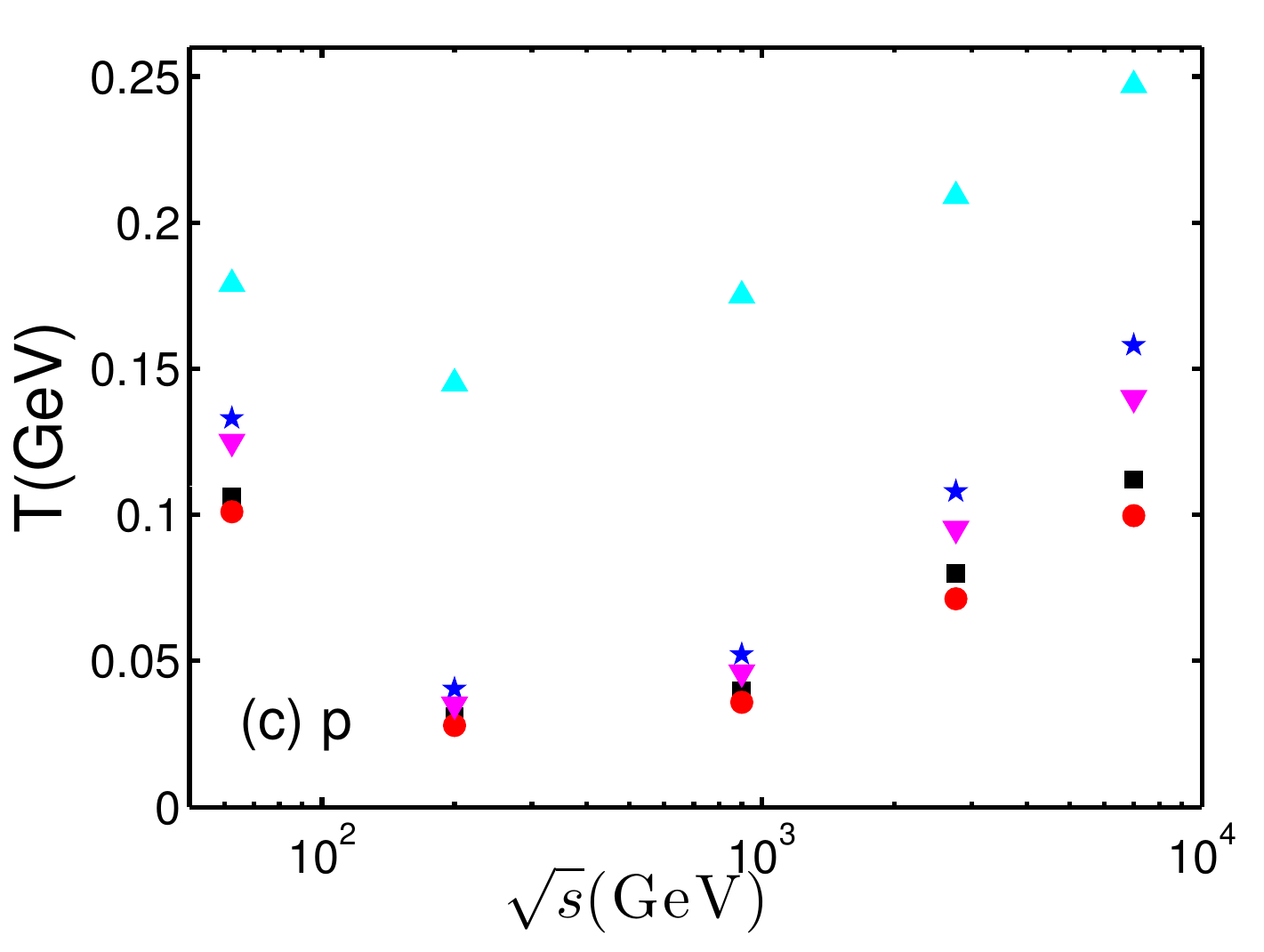}
               \end{tabular}
\caption{(Color online) Temperature $T$ in the five distributions versus $\sqrt{s}$ for (a) $\pi$, (b) K and (c) p in p+p collisions.}\label{Fig5}
    \end{figure*}

\section{Results}
\label{sec:2}

We fit the particle spectra with different $p_T$ ranges and different rapidity cuts from p+p collisions at $\sqrt{s} = 62.4, 200, 900, 2760$ and $7000$ GeV with the five distributions discussed in the last section. The fitting process is the same as that in refs. {\cite{huapp, huaaa}}. Compared with ref. {\cite{huapp}}, the identified particle spectra data at $\sqrt{s}=2760$ and $7000$ GeV have been updated.

In this work, we are interested in the differences of the parameter $T$ from the five distributions. As we argued in refs. \cite{huapp, huaaa}, $T$ is one free fitting parameter and can be different for different particles even they are produced in the same colliding system. In order to distinguish the parameter $T$ in the five distributions and make our discussion clear, we assign $T_1$, $T_2$, $T_3$, $T_4$ and $T_5$ to Eqs. (\ref{tsallisBR}), (\ref{transeq1}), (\ref{transeq2}), (\ref{transeq3}) and (\ref{tsallisus}), respectively. All the five distributions can fit the particle transverse momentum spectra very well. The values of $T_i$ ($i=1, 2, 3, 4, 5$) and corresponding $\chi_i^2/ndf$ for $\pi$, $K$, $p$ and charged particles are shown in TABLE \ref{fitpara}. Here, we only show the fitting results with the five distributions for four cases:  (1) $\pi^+$ at $\sqrt{s}=200$ GeV, (2) $\pi^+$ at $\sqrt{s}=900$ GeV, (3) $K^++K^-$ at $\sqrt{s}=2760$ GeV, (4) $p+\bar p$ at $\sqrt{s}=7000$ GeV.

 In Figs. \ref{Fig1} and \ref{Fig2}, we show the fitting results for $\pi^+$ at $\sqrt{s}=200$ GeV from STAR Collaboration and  $\sqrt{s}=900$ GeV from ALICE Collaboration using the five distributions Eq. (\ref{tsallisBR}) (solid line), Eq. (\ref{transeq1}) (dashed line), Eq. (\ref{transeq2}) (dotted line), Eq. (\ref{transeq3}) (dash-dotted line) and Eq. (\ref{tsallisus}) (long dash-dotted line), respectively. Since the lines are so close to each other, they are almost indistinguishable in the figures. To visualize the fitting quality better, we also plot the ratios of experimental data and fitting results at the bottom of the figures. We can see that the five distributions can describe the experimentally measured $\pi^+$ transverse momentum spectra very well. The errors are within 20\%.

To show the fitting results other than pions, we select the $K^++K^-$ at $\sqrt{s}=2760$ GeV and $p+\bar p$ at $\sqrt{s}=7000$ GeV at LHC in Figs. \ref{Fig3} and \ref{Fig4}, respectively. For the five distributions, we can not distinguish them at all in the two cases. Similar to pions, the fitting qualities are also very good.

\begin{table*}[hbt]
  \centering
      \caption{The fitting parameter $T$ and corresponding $\chi^2$/ndf in the distributions,  Eqs. (\ref{tsallisBR}, \ref{transeq1}, \ref{transeq2}, \ref{transeq3}, \ref{tsallisus}), for the particle spectra in p+p collisions. The unit of $T$ is GeV.}
 \resizebox{0.9\textwidth}{!}{%
            \begin{tabular}{c|c|c|c|c|c|c|c|c|c|c|c|c}
\hline
data source                                 & $\sqrt{s}$(GeV) & particle & $T_1$    & $\chi_1^2$/ndf     & $T_2$ & $\chi_2^2$/ndf & $T_3$ & $\chi_3^2$/ndf &$T_4$ & $\chi_4^2$/ndf & $T_5$  & $\chi_5^2$/ndf   \\
\hline
PHENIX\cite{phenix2009} &    62.4        & $\pi^0$   &  0.104  &    1.433/11   & 0.097 & 1.451/11  &   0.136  & 1.663/11  &0.126 &  1.656/11& 0.135 & 1.686/11 \\
\hline
PHENIX\cite{phenix2011} &    62.4        & $\pi^+$   &  0.0927   &  6.857/23     & 0.085 & 6.866/23 & 0.133     & 4.767/23  & 0.123 & 4.784/23 & 0.132 & 4.779/23 \\
                 			      &                   & $\pi^- $   &   0.0898  &  8.049/23    & 0.0824 & 8.045/23 & 0.128     &  5.173/23 & 0.118 & 5.198/23 & 0.130 & 5.194/23 \\            		                                               &                   & $K^+$    &   0.0856   &   4.837/13   & 0.0775 & 4.822/13 & 0.122    &  5.141/13  & 0.105 & 5.349/13 & 0.160 & 5.121/13\\
                                                                                    &                   & $K^-$     &   0.0936   &   2.002/13   & 0.0851& 2.006/13 & 0.130   &  2.199/13   & 0.119 & 2.203/13  & 0.163 & 2.186/13\\
                                          &                   & $p$        &    0.106    &   7.017/ 24  & 0.101 & 7.075/24 &  0.133  & 6.934/24  & 0.125 & 6.945/24 & 0.179 & 6.966/24 \\
                                          &                   & $\bar p$ &    0.0635  &   6.605/22  & 0.0588 & 6.563/22 &  0.0831 & 6.037/22     & 0.0817 & 5.079/22 & 0.148 & 7.178/22 \\
\hline
PHENIX\cite{phenix2007} &    200     &   $\pi^0$   & 0.0873 & 18.396/22    & 0.0787 &    18.386/22 & 0.117  &  22.004/22   & 0.105 & 22.946/22 & 0.118  &  23.025/22\\
\hline
PHENIX\cite{phenix2011} &    200     &  $\pi^+$   &  0.0741  &  5.278/24   &  0.0657 & 5.275/24  &  0.111  &  4.491/24  &  0.0981 & 4.515/24 & 0.114  & 4.485/24 \\
                                         &               & $\pi^-$     &  0.0811  &  4.710/24    & 0.0725 & 4.703/24 &  0.121 &  3.350/24 & 0.108 & 3.372/24 & 0.123 & 3.354/24 \\
                                        &               & $K^+$     &  0.0473  &  1.561/13    &  0.0418 & 1.591/13 & 0.0729 & 1.634/13  & 0.0601 & 1.602/13 & 0.138  & 1.587/13\\
                                         &              & $K^-$      &0.0621   &    3.013/13    &  0.0542 &  3.010/13 & 0.0913   &   3.004/13  &  0.0781 & 2.999/13 & 0.147  & 2.999/13 \\
                                        &               & $p$         & 0.0311   &   23.832/31  &  0.0279& 23.659/31 & 0.0404  & 24.004/31 & 0.0350 & 24.272/31 & 0.145 &  24.581/31\\
                                        &                 & $\bar p$  & 0.0473 &    12.902/31  &  0.0426 & 12.970/31 & 0.0609 & 13.240/31 & 0.0547 & 13.153/31 & 0.154  & 13.535/31\\
\hline
STAR\cite{star2} &    200     &   $\pi^+$   &  0.0895 & 6.545/20 &  0.0809 & 6.539/20 &    0.126 &  5.032/20  & 0.113 & 5.008/20 & 0.128 & 5.009/20\\
                           &                & $\pi^-$     &  0.0900  &     6.855/20  & 0.0814 & 6.854/20 &  0.127  &  4.700/20 & 0.114 & 4.718/20 & 0.128 & 4.705/20 \\
                           &                 & $p$         & 0.0804   &     10.683/17  & 0.0735 & 10.653/17 & 0.104  & 10.375/17 & 0.0950 & 10.396& 0.180  & 10.359/17\\
                                            &                 & $\bar p$  & 0.0765 &   10.380/17    & 0.0695 & 10.318/17 & 0.0995 & 10.079/17 & 0.0901 & 10.076/17 & 0.177 & 9.991/17\\
 \hline
ALICE\cite{alice1}   &  900   & $\pi^0$  &  0.0922  &7.991/10&  0.0816  & 8.041/10 & 0.136 & 7.554/10  & 0.119 & 7.551/10 & 0.137 & 7.537/10 \\
\hline
ALICE\cite{alice2}    &   900  & $\pi^+$ & 0.0716 &  24.640/30  & 0.0627&  25.530/30 &  0.123 &  13.528/30 & 0.107 & 13.749/30 & 0.125 & 13.460/30           \\
                                     &          & $\pi^-$   & 0.0727 &   17.138/30 & 0.0636 & 17.602/30 &  0.125  & 12.394/30 & 0.109 & 12.645/30 & 0.126 & 12.483/30        \\
                                     &          & $K^+$    & 0.0568 & 12.790/24 & 0.0488 & 12.807/24  & 0.0904 & 13.034/24 & 0.0749 & 13.069/24 & 0.159 & 12.980/24\\
                                   &           & $K^-$    &   0.0624 & 6.457/24 &   0.0538   & 6.552/24 & 0.0968 & 6.641/24 & 0.0820 & 6.636/24  & 0.161 & 6.609/24\\
                                    &          & $p$    &0.0397 & 13.879/21 & 0.0358 & 13.908/21 & 0.0522 &13.816/21 & 0.0460 & 13.849/21 & 0.175 & 13.974/21\\
                                    &          & $\bar p$& 0.0649 & 13.586/21 &0.0568 &13.674/21 & 0.119 & 14.860/21 & 0.0769 &13.544/21 & 0.188 &13.675/21  \\
\hline
ALICE\cite{alice3}   &  2760   & $\pi^0$  & 0.0951  & 6.164/15 & 0.0835 &  6.162/15  & 0.141 &   5.981/15 &  0.122 & 5.979/15 & 0.142 & 5.987/15   \\
 \hline
ALICE\cite{alice4}    &   2760  & $\pi^++\pi^-$ & 0.0810 & 149.545/ 60 &   0.0706 & 149.71/60 &  0.128 & 31.492/60 & 0.110 &   31.583/60   & 0.130  & 31.583/60      \\
                               &      & $K^++K^-$    & 0.0953 & 10.339/55  & 0.0834 & 10.347/55 & 0.140 & 14.654/55 & 0.120 & 14.655/55  & 0.190 & 14.481/55\\
                                &     & $p+\bar p$    & 0.080 & 30.413/46  &   0.0712 &30.647/46  &0.108 &  33.068/46   & 0.0952 & 32.652/46 & 0.209 & 32.877/46 \\
\hline
ALICE\cite{alice1}   &   7000  & $\pi^0$  &0.0942 & 11.071/30 &  0.0822    & 11.112/30 &  0.142   & 16.664/30 & 0.122 & 16.707/30 & 0.142 &16.750/30           \\
\hline
ALICE\cite{alice5}    &   7000  & $\pi^++\pi^-$   &0.0691&    68.408/38   &0.0587  &  68.193/38   & 0.124 &26.041/38 &   0.104 & 26.539/38 & 0.127 & 26.216/38 \\
                                    &          & $K^++K^-$    &  0.100 &   7.807/45 &  0.0867 &   7.910/45 & 0.158 &   5.335/45  & 0.134   & 5.373/45 & 0.209 & 5.333/45 \\
                                    &          & $p+\bar p$    & 0.112 &  14.906/43 &0.0997 & 14.763/43  & 0.158 & 12.719/43 & 0.140 & 12.684/43 &0.247 & 12.700/43 \\
\hline
CMS\cite{cms1}   &  900           & charged   & 0.0899 & 63.863/17 &  0.0794&  63.940/17 & 0.127 & 85.844/17  & 0.110 &  85.2927/17 &0.129 & 84.867/17  \\
CMS\cite{cms2}   &  2760         & charged   &  0.0937  & 83.860/19 &   0.0818 &  84.026/19 & 0.135 & 110.614/19 & 0.115 & 110.770/19 & 0.136 & 110.4/19 \\
CMS\cite{cms1}   &  7000     &     charged    & 0.101&   64.009/24   &0.0880 & 63.955/24 &0.148 & 80.316/24  &0.126 & 80.353/24 & 0.147 &  80.25/24 \\                                      \hline
  \end{tabular}
}
         \label{fitpara}
\end{table*}

From the above results, we can see that the five distributions have the same fitting power to the particle spectra. But as we can see from TABLE. \ref{fitpara}, the parameter $T_i$ are different. It is the main purpose of this work to target what causes the temperature discrepancies among different Tsallis distributions, with and without thermodynamical description. In order to avoid confusion, we emphasis the meaning of each $T_i$. Using the Tsallis distribution classification in ref. \cite{huapp}, $T_1$ refers to Type-B Tsallis distribution and $T_5$ refers to Type-A Tsallis distribution. $T_2$, $T_3$ and $T_4$ are from the transient distributions to bridge the Type-A and Type-B Tsallis distributions. To have a clear picture for the parameter $T_i$ from the five distributions, we plot $T_i$ versus the colliding energy $\sqrt{s}$ for $\pi$, $K$ and $p$, as shown in Fig. {\ref{Fig5}}. We can clearly see that the Type-B Tsallis distribution gives lower $T$ than the Type-A Tsallis distribution, as we mentioned in ref. \cite{huapp}. For all the particles, $T_1$ and $T_2$ which are from the distributions, Eqs. (\ref{tsallisBR}, \ref{transeq1}), with extra $m_T$ term are lower than $T_3$, $T_4$ and $T_5$ which are from the distributions, Eqs. (\ref{transeq2}, \ref{transeq3}, \ref{tsallisus}), without it. As we see that $T_1$ is larger than $T_2$, since Eq. (\ref{tsallisBR}) and Eq. (\ref{transeq1}) are similar except the power in the distributions, we conclude that the power $q$ in Eq. (\ref{tsallisBR}) causes larger $T$. This can be verified by comparing $T_3$ with $T_4$. With the same argument, since $T_4$ is smaller than $T_5$, the $m_T$ in Eq. (\ref{transeq3}) causes smaller $T$. To see the effects of $q$ and $m_T$ in the distribution, we can compare $T_3$ with $T_5$. For the light particles, i.e., pions (left panel), $T_3$ and $T_5$ are similar. The effects of the power $q$ and $m_T$ cancel each other. But for the heavier particles, i.e., kaons and protons (middle and right panels), the effect of $m_T$ in Eq. (\ref{transeq2}) wins and $T_3$ is smaller than $T_5$. We note that the effect of the extra $m_T$ term is crucial for the temperature difference between Type-A and Type-B Tsallis distributions, especially for the light particles. While for the heavier particles, the effect of the choice of $m_T$ or $E_T$ in the Tsallis distribution becomes more important as we can see that $T_5$ is larger than the other $T_i$ for kaons and protons in Fig. \ref{Fig5}.

\section{Summary}
In this paper, we have presented a detailed investigation of two types of Tsallis distribution, with and without the thermodynamical description, by the $p_T$ spectra measured from STAR, PHENIX Collaborations at RHIC and ALICE, CMS Collaborations at LHC. The power $q$ in the Tsallis distribution with thermodynamical description is responsible for the thermodynamical consistency. To show a clear and complete comparison, another three transient distributions to bridge the two types of Tsallis distribution are also given. The five distributions have the similar fitting power to the particle spectra. The good agreements are obtained, but they give different temperatures. Agreed with our previous work \cite{huapp}, the Tsallis distribution with thermodynamical description gives lower $T$ than the ones given by the distribution without it. The extra term $m_T$ in the Tsallis distribution with thermodynamical description is responsible for the discrepancies of the temperatures. But for the heavier particles, the effect of the choice of $m_T$ or $E_T$ in the Tsallis distribution beats the effect of the extra term $m_T$. The data for p+p collisions at 8 and 13 TeV are coming. We wish these data would also support our conclusion presented here.

\section*{Conflict of Interests}
The authors declare that there is no conflict of interests regarding the publication of this article.

\section*{Acknowledgments}
We thank the anonymous referee who brought us to this topic. This work is supported by the NSFC of China under Grant No.\ 11205106.

\end{document}